\documentstyle[aps,epsfig]{revtex}

\begin{document}
\title{Giant vortices in small mesoscopic disks: an approximate description}
\author{S. V. Yampolskii \cite{permadr} and F. M. Peeters \cite{eladr}}
\address{Departement Natuurkunde, Universiteit Antwerpen (UIA),\\
Universiteitsplein 1, B-2610 Antwerpen, Belgium}
\date{\today}
\maketitle

\begin{abstract}
We present an approximate description of the giant vortex state in a thin
mesoscopic superconducting disk within the phenomenological Ginzburg-Landau
approach. Analytical asymptotic expressions for the energies of the states
with fixed vorticity are obtained when a small magnetic flux is accumulated
in the disk. The spectrum of the lowest Landau levels of such a disk is also
discussed.

\end{abstract}

\section{Introduction}

Progress in microfabrication technology has made mesoscopic superconductors
(mesoscopic disks, in particular) a very popular subject of study (see, for
example, Refs.~\cite{geim,palacios}). Mesoscopic samples have sizes
comparable to the coherence length~$\xi $ and the magnetic field penetration
length~$\lambda $. The behaviour of such structures in an external magnetic
field~$H$ is strongly influenced by the sample shape~\cite{shape} and may
lead to various superconducting states~\cite{fink,mv1}.

The vortex structure in mesoscopic disks with finite thickness was studied
early using a numerical solution of the system of two coupled non-linear
Ginzburg-Landau (GL) equations~\cite{numer}. For {\it thin} disks it is
possible to simplify the solution of the problem by averaging the order
parameter over the disk thickness~\cite{mv1} and by representing the
superconducting order parameter as a superposition of eigenfunctions of the
linearized GL equation~\cite{palacios,mv1,analyt1}. But, even then the
problem has still to be solved numerically because of the presence of
confluent hypergeometric functions.

It was shown previously that in some limiting cases the description of the
vortex structure can be simplified. Buzdin and Brisson~\cite{buzdin}
described certain vortex configurations in a small superconducting disk
within the London approximation. Akkermans {\it et al.} calculated the
vortex structure of mesoscopic disks at the vicinity of the dual point $%
\kappa ^{2}=1/2$~\cite{akkerm1} and in the London limit~\cite{akkerm2}.
Recently, the structure of a giant vortex in an infinite plane was analyzed
analytically within the GL theory for arbitrary values of $\kappa $ in the
limits of small and large values of the vorticity~\cite{mallik}.

In the present paper we obtain analytical results for the free energy of
small disks within the approach of Refs.~\cite{mv1,analyt1}, using the
magnetic flux trapped by the disk as an expansion parameter.

\section{General theoretical formalism}

We consider a mesoscopic superconducting disk with radius $R$ and thickness $%
d<<\lambda ,\xi $ magnetized by the external magnetic field $\vec{%
H}=(0,0,H)$ which is uniform and directed normal to the disk plane. The
theoretical model was already described in detail in Refs.~\cite{mv1,analyt1}
and therefore we scetch only those steps which are necessary for our
analytical approach. For a thin disk, to a first approximation, the magnetic
field is uniform inside the disk and equal to the external one. As a result
of this approximation, the distribution of the superconducting order
parameter in the disk plane $\psi (\vec{\rho })$ is described by
the first GL equation 
\begin{equation}
\left( -i\vec{\nabla }_{2D}-\vec{A }\right) ^{2}\psi
=\psi \left( 1-\left| \psi \right| ^{2}\right) ,  \label{GL1}
\end{equation}
with $\vec{A }(\rho )=(0,H\rho /2,0)$ and with the boundary
condition at the sample surface 
\begin{equation}
\left. \left( -i\vec{\nabla }_{2D}-\vec{A }\right) \psi
\right| _{\rho =R}=0.  \label{BC1}
\end{equation}
The index $2D$ refers to the two-dimensional operator. Due to the circular
symmetry of the sample we use cylindrical coordinates: $\vec{\rho 
}=(\rho ,\theta )$ ($\rho $ is the radial distance from the disk center, $%
\theta $ is the azimuthal angle). All distances are measured in units of the
coherence length~$\xi $, the magnetic field in $H_{c2}=2^{1/2}\kappa H_{c}$,
where $H_{c}$ is the thermodynamical critical field.

In the giant vortex state~\cite{fink} the order parameter can be expressed
as 
\begin{equation}
\psi \left( \rho ,\theta \right) =\left( -\Lambda \frac{I_{1}}{I_{2}}\right)
^{1/2}f_{L}\left( \rho \right) \exp \left( iL\theta \right) ,  \label{OP1}
\end{equation}
where 
\begin{equation}
f_{L}\left( \rho \right) =\left( \frac{H\rho ^{2}}{2}\right) ^{L/2}\exp
\left( -\frac{H\rho ^{2}}{4}\right) M\left( -\nu ,L+1,\frac{H\rho ^{2}}{2}%
\right) ,  \label{eigF1}
\end{equation}
\begin{equation}
I_{m}(L,\Phi )=\int\limits_{0}^{\Phi }t^{mL}\exp \left( -mt\right) \left[
M\left( -\nu ,L+1,t\right) \right] ^{2m}dt,  \label{InteigF1}
\end{equation}
are the eigenfunctions of the linearized Eq.~(\ref{GL1}) and 
\begin{equation}
\Lambda =-1+\frac{2\Phi }{R^{2}}\left( 1+2\nu \right)  \label{eigenval1}
\end{equation}
determines the spectrum of the lowest Landau levels, 
\begin{equation}
M\left( a,c,z\right) =1+\sum\limits_{k=1}^{\infty }\frac{\Gamma \left(
a+k\right) \Gamma \left( c\right) }{\Gamma \left( a\right) \Gamma \left(
c+k\right) }\frac{z^{k}}{k!}  \label{Kummer1}
\end{equation}
is the Kummer function~\cite{book}. The value of $\nu $ is determined by a
non-linear equation, which results from the boundary condition~(\ref{BC1}) 
\begin{equation}
\left( L-\Phi \right) M\left( -\nu ,L+1,\Phi \right) -\frac{2\nu \Phi }{L+1}%
M\left( -\nu +1,L+2,\Phi \right) =0,  \label{equatBC}
\end{equation}
where $\Phi =HR^{2}/2$ (measured in units of $\Phi _{0}=\pi \hbar c/e$) is
the magnetic flux through the disk in the absence of any flux expulsion. The
free energy, measured in $F_{0}=H_{c}^{2}V/8\pi $ units, is 
\begin{equation}
F=-\frac{\Lambda ^{2}}{\Phi }\frac{I_{1}^{2}\left( L\right) }{I_{2}\left(
L\right) }.  \label{free_en1}
\end{equation}

\section{Approximate results}

First, we have to solve the non-linear Eq.~\ref{equatBC}. In the case of
small $\Phi $ we can restrict the series~(\ref{Kummer1}) for the Kummer
function by the first four terms. The solution of Eq.~\ref{equatBC} can be
written as an infinite series of which the first four terms are 
\begin{equation}
\nu =\frac{\left( L+1\right) \left( L-\Phi \right) }{2\Phi }\left[ 1-\frac{%
L+\Phi }{2\left( L+2\right) }+\frac{L\left( L+\Phi \right) }{2\left(
L+2\right) ^{2}\left( L+3\right) }+\frac{\left( L+\Phi \right) \left( \Phi
^{2}-L^{2}+2L^{3}\right) }{4\left( L+2\right) ^{3}\left( L+3\right) \left(
L+4\right) }\right] +\ldots . \label{njuas1}  
\end{equation}

In the limit of large $\Phi >>1$ we use the asymptotic expression for the
Kummer function 
\begin{equation}
M\left( a,c,z\right) =\frac{\Gamma (c)}{\Gamma (a)}z^{a-c}\exp (z)\left[
1+O\left( \left| z\right| ^{-1}\right) \right] .  \label{Kumasinf}
\end{equation}
Substituting it in Eq.~\ref{equatBC} we obtain the equation $(L+\Phi
)/\Gamma (-\nu )=0$, which leads to the solution 
\begin{equation}
\nu _{\inf }=0.  \label{njuasinf}
\end{equation}

\subsection{Lowest Landau levels}

Substituting $\nu $ from Eq.~(\ref{njuas1}) into Eq.~(\ref{eigenval1}) we
obtain the eigenvalues $\Lambda $ in the $\Phi <<1$ limit: 
\begin{eqnarray}
\Lambda  &=&-1+\frac{2L\left( L+1\right) }{R^{2}}\left[ 1-\frac{L}{2\left(
L+2\right) }+\frac{L^{2}}{2\left( L+2\right) ^{2}\left( L+3\right) }+\frac{%
L^{3}\left( 2L-1\right) }{4\left( L+2\right) ^{3}\left( L+3\right) \left(
L+4\right) }\right]   \label{lam_as} \\
&&-\frac{2L}{R^{2}}\Phi +\frac{\left( L+1\right) }{\left( L+2\right) R^{2}}%
\left[ 1-\frac{L}{\left( L+2\right) \left( L+3\right) }-\frac{L^{2}\left(
L-1\right) }{\left( L+2\right) ^{2}\left( L+3\right) \left( L+4\right) }%
\right] \Phi ^{2}  \nonumber \\
&&-\frac{L+1}{2\left( L+2\right) ^{3}\left( L+3\right) \left( L+4\right)
R^{2}}\Phi ^{4}.  \nonumber
\end{eqnarray}
Because series~(\ref{njuas1}) is asymptotic, each new term gives a
contribution to all the coefficients of the $\Phi $-series.

\begin{figure}[tbp]
\begin{center}
\epsfig{file=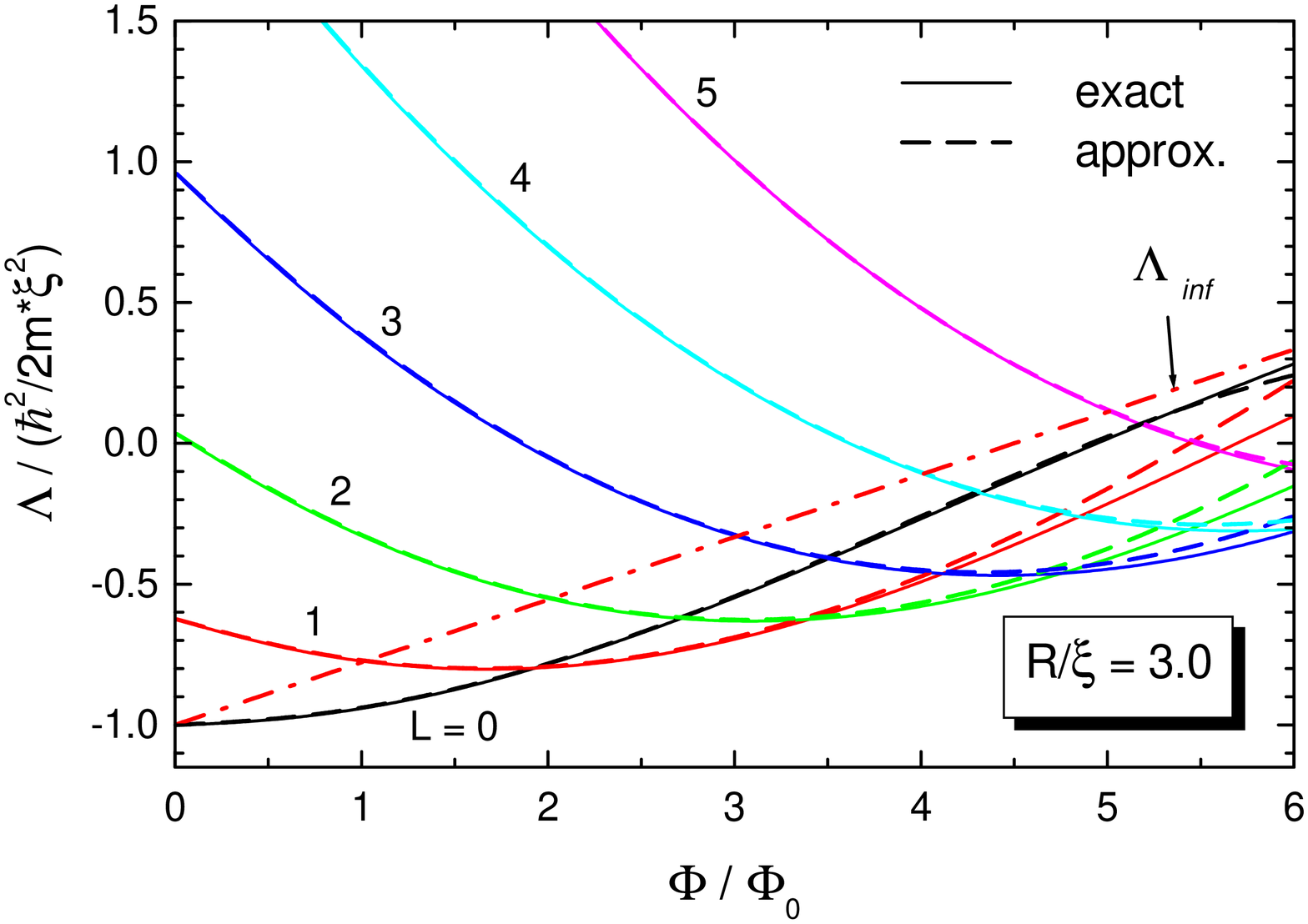, width=120mm, height=90mm, clip=}
\end{center}
\caption{The magnetic flux dependence of the lowest eigenvalues of the
linearized GL equation~(\ref{GL1}) for different vorticities~$L$. The
``exact'' numerical results are shown by the solid curves. The approximate
results are shown by the dashed curves ($\Phi <<1$ limit) and by the
dash-dotted curve ($\Phi >>1$ limit).}
\label{fig1}
\end{figure}

In the $\Phi >>1$ limit we obtain for $\Lambda $ the asymptotic expression 
\begin{equation}
\Lambda _{\inf }=-1+\frac{2\Phi }{R^{2}},  \label{lam_asinf}
\end{equation}
which does not depend on the vorticity $L$. In Fig.~\ref{fig1} the $\Lambda
(\Phi )$ dependences are shown for a disk with radius $R=3\xi $. Only the
curves with $L<6$ are shown. The solid curves represent the dependences
calculated from the numerical solution of expressions~(\ref{eigenval1}) and~(%
\ref{equatBC}) and the dashed curves are our new approximate results as
given by Eq.~(\ref{lam_as}). Notice that for small values of the magnetic
flux the approximate expression~(\ref{lam_as}) describes quite well the
lowest Landau level spectrum.

\subsection{Free energy of giant vortices in the small $\Phi $ limit}

The eigenvalues $\Lambda $, Eq.~(\ref{eigenval1}), determine the minimal
free energy $F$, Eq.~(\ref{free_en1}), of the giant vortices. We expand Eq.~(%
\ref{free_en1}) in a series with respect to $\Phi $. We found that for small
disks with radius $R=(2\div 3)\xi $ it is sufficient to keep in the series~(%
\ref{Kummer1}) only the first four terms. Expanding the free energy up to
order $\Phi ^{4}$ we obtain the approximate expressions: 
\begin{equation}
F_{L=0}=-1+\frac{\Phi ^{2}}{R^{2}}-\left( \frac{557}{2880}+\frac{1}{96R^{2}}+%
\frac{1}{4R^{4}}\right) \Phi ^{4},  \label{FL0}
\end{equation}
\begin{eqnarray}
F_{L=1} &=&-1.16+\frac{7.87}{R^{2}}-\frac{13.34}{R^{4}}+\left( 3.10-\frac{%
25.66}{R^{2}}+\frac{51.38}{R^{4}}\right) \Phi -\left( 18.47-\frac{139.08}{%
R^{2}}+\frac{263.87}{R^{4}}\right) \Phi ^{2}  \label{FL1} \\
&&+\left( 96.56-\frac{732.51}{R^{2}}+\frac{1388.80}{R^{4}}\right) \Phi
^{3}-\left( 501.67-\frac{3810.91}{R^{2}}+\frac{7235.91}{R^{4}}\right) \Phi
^{4}.  \nonumber
\end{eqnarray}
The free energy for the next $L$ states has the same form as $F_{L=1}$ but
with different numeric coefficients and, therefore, will not be give here.

\begin{figure}[tbp]
\begin{center}
\epsfig{file=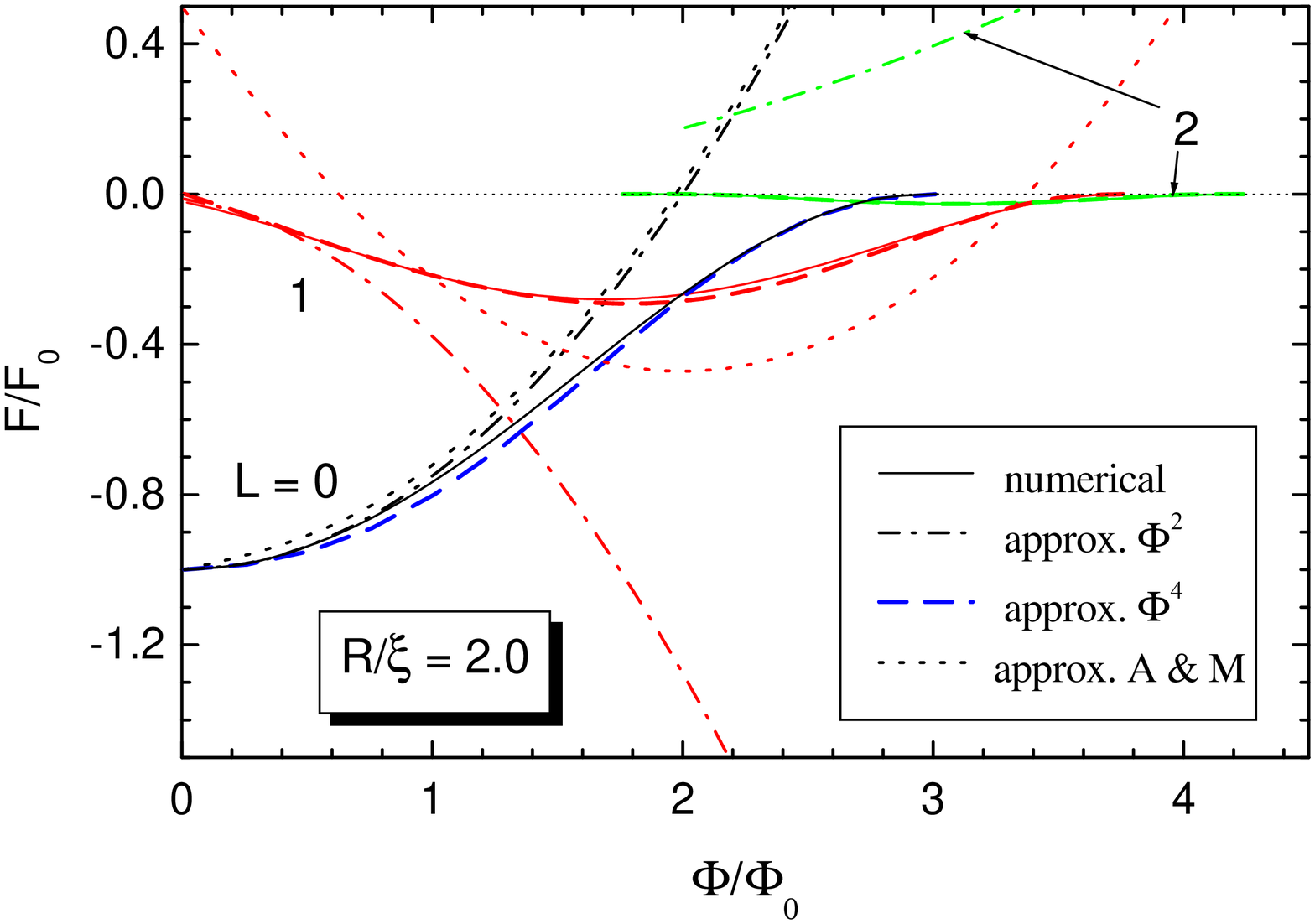, width=120mm, height=90mm, clip=}
\end{center}
\caption{The free energy of the giant vortex states for a thin disk with
radius $R/\protect\xi =2.0$. The solid curves are the numerical solution and
the dashed curves are from our analytical approximation. The dash-dotted
curves are from our analytical expansion up to the $\Phi ^{2}$ term. The
dotted curves are from the analytical expression of Ref.~\protect\cite
{akkerm3}.}
\label{fig2}
\end{figure}

In Fig.~\ref{fig2} the free energy of the disk with $R=2\xi $ is shown.
Notice that limiting oneselves to terms in $F$ up to $\Phi ^{2}$ leads to a
poor approximation, while including terms up to $\Phi ^{4}$ results in an
excellent agreement with the numerical results. The dotted curves in Fig.~%
\ref{fig2} present the free energy calculated with the expression 
\begin{equation}
F(L,\Phi )=-1+L-\frac{\Phi ^{2}}{R^{2}}+\frac{\left( L-\Phi \right) ^{2}}{R}-%
\frac{\left( L-\Phi \right) ^{4}}{2R^{3}},  \label{F_Akkerm}
\end{equation}
proposed in Ref.~\cite{akkerm3} (and rewritten in the notation of the
present paper). Notice that expression~(\ref{F_Akkerm}) describes less
accurately our numerical result (solid curves). It agrees for $L=0$ with our 
$\Phi ^{2}$ approximation, but for $L=1$ there is a substantial deviation
with our approximate result. This disagreement is even larger for $L=2$ and
is therefore not given.

\section{Conclusion}

We calculated approximate expressions for the lowest Landau levels and the
free energy of the giant vortex states in thin mesoscopic disk with small
radius. We found that these approximate analytical expressions agree very
well with the more involved numerical calculation~\cite{mv1,analyt1} and are
a substantial improvement to those found in Ref.~\cite{akkerm3}.

\vspace{5mm} {\bf Acknowledgement. }This work was supported by the Flemish
Science Foundation (FWO-Vl), the ``Onderzoeksraad van de Universiteit
Antwerpen'' (GOA), the ``Inter-University Poles of Attraction Program -
Belgian State, Prime Minister's Office - Federal Office for Scientific,
Technical and Cultural Affairs'', and the ESF ``VORTEX'' Programme.

\end{document}